\documentclass[reprint,showpacs,preprintnumbers,prb,amsmath,amssymb,floatfix]{revtex4-1}
\usepackage{graphicx,epsf,epsfig,epstopdf}
\usepackage{graphicx}
\usepackage{subfigure}

\begin{document}
%
\title{Trap assisted tunneling and its effect on subthreshold swing of tunnel field effect transistors}
\author{Redwan N. Sajjad}
\affiliation{Microsystems Technology Laboratories, Massachusetts Institute of Technology, Cambridge, MA-02139}  

\author{Winston Chern}
\affiliation{Microsystems Technology Laboratories, Massachusetts Institute of Technology, Cambridge, MA-02139}  
\author{Judy L. Hoyt}
\affiliation{Microsystems Technology Laboratories, Massachusetts Institute of Technology, Cambridge, MA-02139}  
\author{Dimitri A. Antoniadis}
\affiliation{Microsystems Technology Laboratories, Massachusetts Institute of Technology, Cambridge, MA-02139}


\begin{abstract}
We provide a detailed study of the interface Trap Assisted Tunneling (TAT) mechanism in tunnel field effect transistors to show how it contributes a major leakage current path before the Band To Band Tunneling (BTBT) is initiated. With a modified Shockley-Read-Hall formalism, we show that at room temperature, the phonon assisted TAT current  always dominates and obscures the steep turn ON of the BTBT current for common densities of traps. Our results are applicable to top gate, double gate and gate all around structures where the traps are positioned between the source-channel tunneling region. Since the TAT has strong dependence on electric field, any effort to increase the BTBT current by enhancing local electric field 
also increases the leakage current. Unless the BTBT current can be increased separately, calculations show that the trap density $D_\mathrm{it}$ has to be decreased by 40-100 times compared with the state of the art in order for the steep turn ON (for III-V materials) to be clearly observable at room temperature. We find that the combination of the intrinsic sharpness of the band edges (Urbach tail) and the surface trap density determines the subthreshold swing. 
\end{abstract}

\maketitle


\section{Introduction}

The Tunnel Field Effect Transistor (TFET) \cite{seabaugh2010} is a candidate for low power switching in digital logic circuits for replacing or supplementing standard CMOS technologies because of its potential to 
reduce power dissipation via reduction of the power supply voltage. 
In a TFET, over-the-barrier thermionic emission is completely bypassed by triggering a BTBT current by the gate voltage, allowing steep ``subthermal" change of current and reduced supply voltage. It has been shown that a small reduction in the subthreshold swing (SS) (e.g. to 45-53 mV/dec) in TFET can reduce the dynamic power dissipation by at least 50\% \cite{avci2013,young2015} with little sacrifice on the switching delay. Such energy saving is calculated for the same OFF current but lower ON current (compared to the CMOS). The energy savings may enable high frequency operation that currently CMOS cannot provide. Further improvement is possible if higher ON current is achieved, which can be done with III-V semiconductors and heterojunctions \cite{knoch2010}.

However, the ideal picture of TFET operation is based upon the assumption that the Band To Band
Tunneling (BTBT) current is sufficiently higher than any background current that flows before the bands overlap. In an ideal TFET operation, very little current should flow for gate voltages below a threshold voltage (defined as the gate voltage when the conduction band bottom in the channel and the valence band extrema in the source first overlap) and a large amount of current should flow above that. Such notion of steep (or ideal) switching is practically difficult to achieve since the combined leakage current, e.g. gate or substrate leakage, bulk or interface trap assisted tunneling will always be present and can easily obscure steep change of the BTBT current near the threshold voltage. In addition, the steepness of the current change partly depends on the BTBT magnitude and since it can be weak for multiple reasons, achieving the steep change of current is highly challenging. Despite numerous efforts in this field, experimental demonstrations with steep turn ON are few \cite{appenzeller_04,choi2007,krishnamohan2008,jeon2010,dewey2011,gandhi2011} and mostly at very low current levels. Most of the demonstrations involved silicon, for which the interface and bulk defect density is by far the best compared to other materials. Except for Refs. \cite{dewey2011,ganjipour2012}, most TFET experimental results on III-V semiconductors \cite{mookerjea2010,tao2015,zhao2014} do not show subthermal switching. On the contrary the subthreshold swing in these experiments shows strong temperature dependence, clearly indicating the existence of a thermal process. Two dimensional layered heterostructure based TFETs have attracted significant attention in recent times with one experiment \cite{sarkar2015} showing low subthreshold swing. But similar structures by other groups \cite{roy2015,amir2016} have failed to produce such behavior.

\begin{figure}
\centering
\includegraphics[width=2.5in]{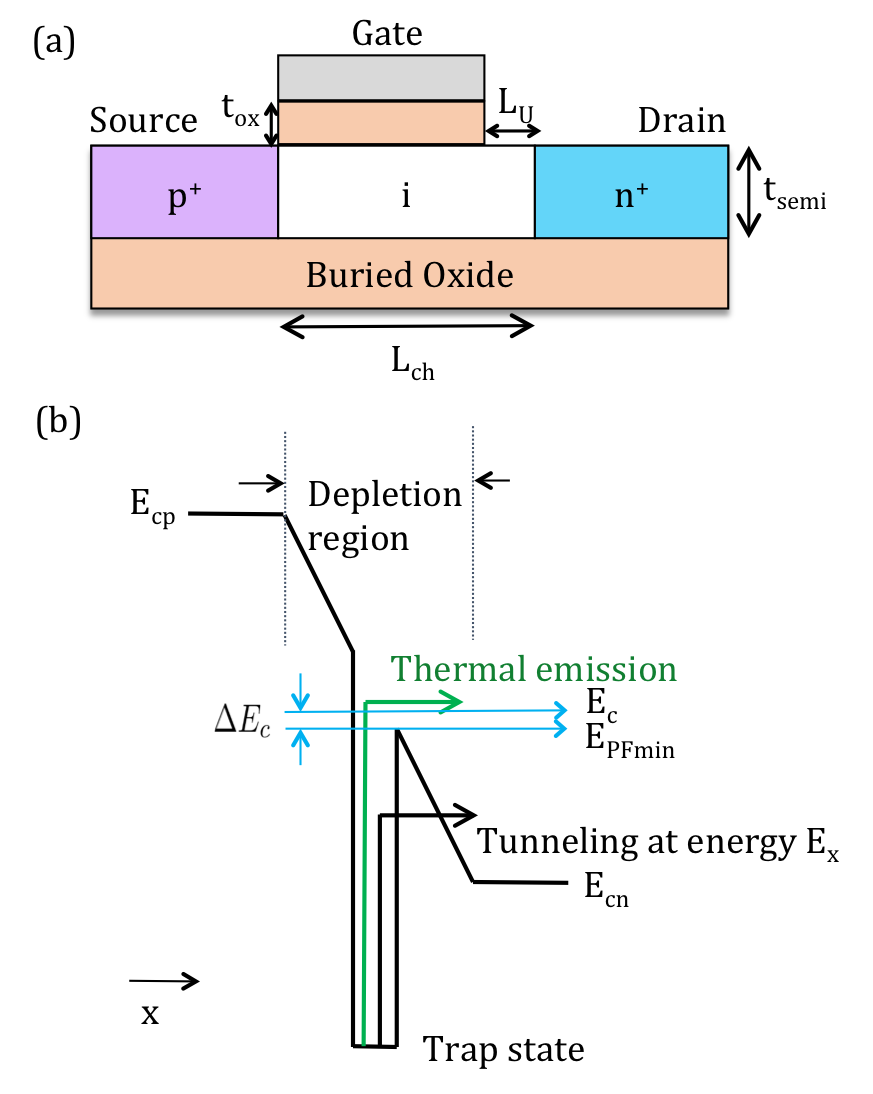}
\caption{(a) Schematic of the top gate device considered in this paper.
(b) Schematic of the trap assisted tunneling (TAT) process: an electron can reach the conduction band from the valence band via a combination of phonon absorption and tunneling. This undesired tunneling is electric field dependent in the same way as the ON state BTBT current. The electric field enhanced generation rate is much higher than the classical Shockley-Read-Hall (SRH) formalism that does not take the electric field into account.
}
\label{fig:device}
\end{figure}

In this work, we show that interface trap assisted tunneling (TAT) current, which is known as a leakage current mechanism in conventional $pn$ junction diodes \cite{schenk1992,hurkx1992}, is also a major parasitic current component in TFETs.  
TAT is the emission of electrons to a trap state via electron-phonon interaction, followed by tunneling into the conduction band (Fig. \ref{fig:device}). Similarly a hole emission and tunneling from a trap is possible. This process is strongly temperature dependent compared to other non-idealities such as exponential band tails from the heavy source doping \cite{khayer2011}. Such interband transition is also possible when phonon scattering is considered alone. Models with phonon scattering (without traps) have shown higher OFF current without sacrificing much on the subthreshold swing \cite{koswatta2008}. Although TAT has been identified in the past as a leakage mechanism in TFETs \cite{mookerjea2010,vallett2010,pala2013,qiu2014,avci2015,sapan2015}, a detailed quantitative study of its deleterious effects has not been performed. We show that in the presence of traps, electron capture rate prescribed by the Shockley-Read-Hall (SRH) formalism is greatly enhanced due to the high electric field near the source. This is due to the fact that the undesirable electron tunneling from trap to conduction band  depends on the local electric field (Fig. 2), in much the same way as the ON state BTBT current. We show that at room temperature this TAT current overshadows the steepest part of the 
BTBT current (Fig. 3) for realistic trap density (midgap $D_\mathrm{it} = 5\times10^{12}$/cm$^{2}$-eV for III-V). The steep turn ON of the BTBT current is observable at low temperatures, where the BTBT dominates and the subthreshold swing becomes less temperature dependent. In our model, we consider the Poole-Frenkel  effect \cite{furlan2001} - the lowering of the electron barrier due to the Coulomb interaction of the trap with the lattice. We find that the Poole-Frenkel effect causes a substantial increase in the leakage current by enhancing the trap-channel tunneling. 
In the next section, we review the electric field dependent SRH formalism, the electrostatic model and the BTBT model used in this paper followed by discussions. The formalism is also applicable to most other device geometries and materials provided that the $D_\mathrm{it}$ is known and the electrostatic configuration is solved appropriately. 
\section{Model description}
In this section, we review the electric field enhanced carrier generation rate via phonon and trap states. We incorporate the Poole-Frenkel effect and the tunneling enhanced rates as done in Refs. \cite{hurkx1992,furlan2001} but apply them using the TFET electrostatics and consider only the surface trap states. The electric field profile from the electrostatic model is used in calculating both the TAT and BTBT current.
\subsection{Trap assisted tunneling} The classical SRH formalism \cite{sze2006} describes the generation rate of electron and hole pairs in presence of traps. An electron in the valence band (VB) can absorb a phonon to reach a trap state before emitting to the conduction band (CB) by interaction with another phonon. However, in the presence of electric field, the trap-CB (or VB-trap) tunneling rate becomes substantial and greatly increases the electron-hole generation rate \cite{hurkx1992}. The net generation rate (per unit area) at a given position in the $pn$ junction space charge region becomes,
\begin{eqnarray}\label{eq:srh}
G^n = \int\frac{n_i^2-np}{\tau_p\frac{n+n_1}{1+\Gamma_p}+\tau_n\frac{p+p_1}{1+\Gamma_n}}D_\mathrm{it}\,dE
\end{eqnarray}where $n_i$ is the intrinsic carrier concentration, $n$ and $p$ are the electron and hole densities, $\tau$ is the minority carrier lifetime and $\Gamma$ is a factor that accounts for the tunneling from trap to CB. With $\Gamma=0$, Eq. \ref{eq:srh} reduces to the classical SRH formalism. The carrier lifetime depends upon the capture cross section and the thermal velocity, $\tau = 1/(\sigma v_{th})$. $\Gamma$ is electric-field-dependent and it effectively decreases the minority carrier lifetime. When the electric field is weak, $\Gamma$ is negligible and Eq. \ref{eq:srh} reduces to the classical SRH formalism. The terms $n_1$ and $p_1$ arise from the principle of detailed balance \cite{sze2006} and are given by $n_1 = n_i\mathrm{exp}^{(E_t-E_i)/k_BT}$, $p_1 = n_i\mathrm{exp}^{(E_i-E_t)/k_BT}$, where $E_i$ and $E_t$ are the position of the Fermi level for intrinsic semiconductor and the trap state. Fig. 1b schematically shows the TAT, which is a two-step process. In the first step, the electron is emitted from the valence band to the trap state by absorbing a phonon. Afterwards, the electron can be partially lifted further and then tunnel into the conduction band. The amount of the partial lift in the second step can vary from $E_{cn}$, the position of the CB in the channel, to $E_c$ the position of the CB at the position under consideration. Within the energy range $E_\mathrm{PF}<E<E_c$, electrons reach the conduction band without any resistance since there is no barrier to tunnel, whereas for $E_{cn}<E<E_\mathrm{PF}$, the transmission probability $T(E)$ through the barrier has to be accounted for. Therefore the enhancement factor $\Gamma$ is the sum of two components, one for each energy regime
\begin{eqnarray}\label{gamma}
\Gamma = \Gamma_\mathrm{PF}+\Gamma_\mathrm{tunnel}
\end{eqnarray} $\Gamma$ is calculated from the net flux (carrier density times the thermal velocity) and the transmission probability \cite{furlan2001},
\begin{eqnarray}\label{eq:gamma}
\Gamma = \frac{1}{k_BT}\int\mathrm{exp}(\frac{E_c-E_x}{k_BT})T(E_x)dE_x
\end{eqnarray}$E_x$ is the energy to which the electron (or hole) is tunneling to (Fig. \ref{fig:device}b). $T(E_x)$ is calculated for a triangular barrier using the Wentzel-Kramers-Brillouin (WKB) approximation. 
\begin{eqnarray}
T(E_x) = \mathrm{exp}(-\frac{4\sqrt{2m^*(E_\mathrm{PF}-E_x)^3}}{3q\hbar F})
\end{eqnarray}where $F$ is the electric field at a particular position in the depletion regime for a given gate voltage. For $E_\mathrm{PF}<E<E_c$, $T(E_x) = 1$. From Eq. \ref{eq:gamma} it can be shown, 
\begin{eqnarray}\label{eq:gamma2}
\Gamma_\mathrm{tunnel} &=& \frac{\Delta E_{n,p}}{k_BT}\int^1_0\mathrm{exp}[\frac{\Delta E_{n,p}}{k_BT}u-K_{n,p}u^{3/2}]\mathrm{du}\\
\Gamma_\mathrm{PF} &=& \frac{1}{4}\mathrm{exp}(\frac{E_c-E_\mathrm{PF}}{k_BT})\nonumber\\
K_{n,p} &=& \frac{4}{3}\frac{\sqrt{2m^*\Delta E_{n,p}^3}}{q\hbar F}\nonumber
\end{eqnarray}where $\Delta E_c = E_c-E_\mathrm{PF}$ is the lowering of the barrier (Fig. 1b) due to the Poole-Frenkel effect. $\Delta E_{n,p}$ is effectively the tunnel barrier height and it also defines the range of energy to which the electron (or hole) can tunnel to (from the trap). So $\Delta E_{n,p}$ is the difference between the top of the barrier and the minimum energy where the electron can tunnel to. Depending upon the position (in the depletion region) under consideration, this can vary from $\Delta E_{n,p} = E_\mathrm{PF}-E_t$ (if $E_t>E_{cn}$) to $E_\mathrm{PF}-E_{cn}$ (if $E_t<E_{cn}$) \cite{hurkx1992}. The higher the Poole-Frenkel effect, the higher the $\Delta E_c$ and the higher the $\Gamma$'s in Eq. \ref{eq:gamma2}. For typical electric fields, the second term in Eq. \ref{eq:gamma2} (which signifies the tunneling contribution), dominates over the first term and increases the exponential term for smaller $\Delta E_{n,p}$ or larger $F$. The lowering of the energy barrier $\Delta E_c$ is determined by the electric field \cite{woo1987,pelaz1994,huang1997} $\Delta E_c = q\sqrt{qF/(\pi\epsilon)}$, where $\epsilon$ is the electric permittivity. The $\Gamma$'s are calculated for both electron and hole (so that all combinations of phonon absorption and tunneling as shown in Fig. \ref{fig:device}b are included in the model) and used in Eq. \ref{eq:srh}.  

Performance degradation in TFET can take place even without the traps due to inelastic phonon scattering \cite{koswatta2008,yoon2012}. The OFF current is increased in addition to making the transfer I-V ambipolar. But the phonon limited subthreshold swing can still be less than 60 mV/dec. Traps on the other hand increase the carrier capture rates to a large extent so that the leakage current dominates over the desired current.  TAT affects both the ON-OFF current ratio and the subthreshold swing. 
\begin{figure}
\centering
\includegraphics[width=3in]{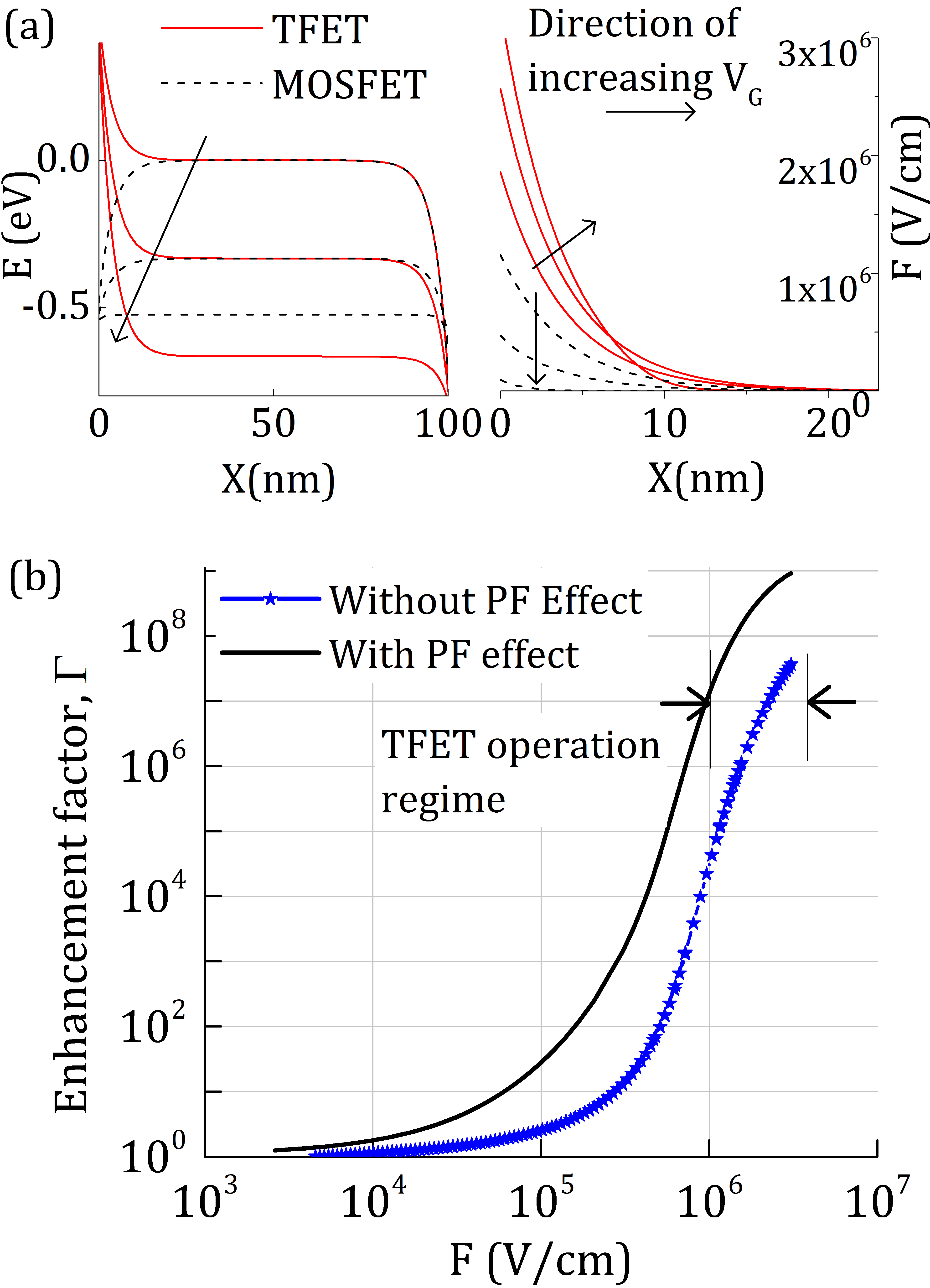}
\caption{Relationship of trap assisted tunneling with electric field ($F$) in tunnel FET. (a) Conduction band profile and the corresponding electric field for silicon TFET at various gate voltages. Solid lines are for TFET and the dotted lines are for a MOSFET configuration. The electric field is increased in TFET as much as possible with gate voltage to increase the band to band tunneling, but in the process it also increases the undesired trap to conduction band tunneling. For the MOSFET on the other hand, the electric field is reduced with gate voltage, taking the trap effects out of the
picture. (b) Carrier lifetime is decreased as a result of trap assisted tunneling by a factor 1+$\Gamma$. $\Gamma$ is large for the typical electric fields in TFETs and increases the generation rate in the source-channel $pn$ junction. Here, $\Gamma$ vs. $F$ is shown at the beginning of the channel ($x=0$).}
\label{fig:model}
\end{figure}

Fig. 2b shows the total enhancement $\Gamma$ (Eq. \ref{gamma}) in silicon with and without the Poole-Frenkel effect. $\Gamma$ can be as high as $10^8$, which is effectively the enhancement of the SRH rate. Typical TFET electric fields operate around $1-5\times10^{6}$ V/cm, over which the $\Gamma$ changes by less than two orders of magnitude. 

Finally the current is calculated from,
\begin{eqnarray}
I/W = q\int G^n(x)\,{dx}
\end{eqnarray}
\subsection{Electrostatic model}
As derived in \cite{yan1992}, we use an abridged version of the 2D Poisson equation for the top gate structure shown in Fig. \ref{fig:device}a. For an SOI structure, the electric field at the top and bottom surface of the semiconductor (given by the oxide thickness and gate potentials) can be applied to the 2D Poisson equation and can be simplified as 
\begin{eqnarray}\label{eq:potential}
\frac{d^2\psi}{dx^2}-\frac{\psi-\phi_{gs}}{\lambda^2} = -\frac{\rho}{\epsilon}
\end{eqnarray} where $\psi$ is the surface potential and $\phi_{gs} = V_G-V_{FB}$ is the gate potential. Eq. \ref{eq:potential} captures the 2D electrostatics quite well for a given characteristic length $\lambda$. For the top gated architecture, $\lambda = \sqrt{\frac{\epsilon_\mathrm{semi}}{\epsilon_\mathrm{ox}}t_\mathrm{ox}t_\mathrm{semi}}$. The charge density in the channel is mainly dictated by the drain injection, since the channel is poorly coupled to the source 
\begin{eqnarray}\label{eq:charge}
\rho \approx -q\,n_0\mathrm{exp}^{{(\psi-V_{DS})}/{k_BT}}+qp_0\mathrm{exp}^{{-\psi}/{k_BT}}
\end{eqnarray} where $n_0$ and $p_0$ are the equilibrium electron concentration in the channel. Eqs. \ref{eq:potential} and \ref{eq:charge} are solved iteratively until self-consistency is achieved. For a given $\rho$, the potential $\psi$ is calculated numerically from Eq. \ref{eq:potential} using the finite difference method subject to appropriate boundary conditions (for the doped regions). Eq. \ref{eq:potential} is also valid for double-gate and gate-all-around nanowire structure if the characteristic length $\lambda$ is changed appropriately \cite{yan1992}. 

Fig. \ref{fig:model}a (left) shows the conduction band profile. On the right, we show the electric field for various gate voltages. For the TFET configuration, the electric field near the source end is greatly enhanced. For a MOSFET configuration on the other hand, the energy barrier (and the conduction band) is pushed down resulting in a decreased electric field near the source. This opposite trend in the electric field with gate voltage, results in a drastically different TAT current in TFET compared to MOSFET  since the TAT is dependent on the local electric field. The TAT for TFETs increases with gate voltage, while for MOSFETs it diminishes quickly (not shown). Therefore the role of traps in MOSFETs is mostly limited to a decreased gate efficiency, while for TFETs, it affects both the gate efficiency and leakage. In this paper, we did not account for the impact of reduced gate efficiency due to $D_\mathrm{it}$. Including it would only increase the subthreshold swing relative to what we show (depending on trap density and gate dielectric thickness).
\subsection{BTBT model} The transmission probability through the tunnel barrier is determined by the WKB approximation \cite{seabaugh2010}. It can be written as,
\begin{eqnarray}
J_\mathrm{wkb} = aV_{TW}(\frac{F}{F_0})^P\mathrm{exp}(-\frac{b}{F})
\end{eqnarray}where $a$, $F_0$, $P$ and $b$ are material parameters taken from \cite{kao2012} and \cite{lu2015}. $V_\mathrm{TW}$ is the tunnel window, i.e.  the energy difference between the valence band in the source and the conduction band in the channel; it is determined by an Urbach tail below the threshold voltage and it increases linearly with gate voltage above the threshold voltage \cite{lu2015}. $V_{TW} = E_0\log\Big[1+\mathrm{exp}\Big(\frac{E_\mathrm{v, source}-E_\mathrm{c,channel}}{E_0}\Big)\Big]$. A difference between Ref. \cite{lu2015} and our approach is that we find the position of the conduction band after self-consistency is achieved between carrier density and channel potential, as discussed in previous sub-section. So for any given gate voltage, the position of the conduction band is $E_\mathrm{c,channel} (\mathrm{V}_G) = E_\mathrm{c,channel}(\mathrm{V_G=0})-\psi$. $E_0$ is the Urbach parameter and it determines the intrinsic subthreshold swing. 

The Urbach tail has been studied in the past in order to understand the sharpness of the optical absorption spectrum in semiconductors. Instead of a steep rise in the absorption co-efficient above a threshold photon energy, experimental results typically show an exponential rise following $\alpha = \alpha_0\mathrm{exp}\Big[-\frac{E-E_g}{E_0}\Big]$\cite{pankove1965,urbach1953}. Such non-abrupt absorption has been attributed to the Urbach tail which originates in heavily doped semiconductors from the  smearing of the dopant energy levels. It can also happen in undoped semiconductors due to electron-phonon interaction \cite{subashiev2010} with a lower Urbach parameter $E_0$. The temperature variation of $E_0$ is weak in doped semiconductors compared to an undoped one \cite{johnson1995}. Unfortunately the exact nature of the Urbach tail and its temperature dependence of $E_0$ is not well understood \cite{greeff1995}. In the next section, we will discuss the implication of various cases of Urbach tail and how it affects the TFET performance.

For a given gate voltage $V_G$, we solve as discussed for the self-consistent channel potential (Eq. \ref{eq:potential}) for the top surface $\psi(x)$ and the electric field $F(x) = -\frac{d\psi}{dx}$. Using the spatial electric field $F(x)$, we calculate the enhancement factors $\Gamma$ from Eq. \ref{eq:gamma}, carrier densities from Eq. \ref{eq:charge} and $x$-dependent generation rate $G^n(x)$ from Eq. \ref{eq:srh}. 
\section{Results and discussion}
\begin{figure}
\centering
\includegraphics[width=3.4in]{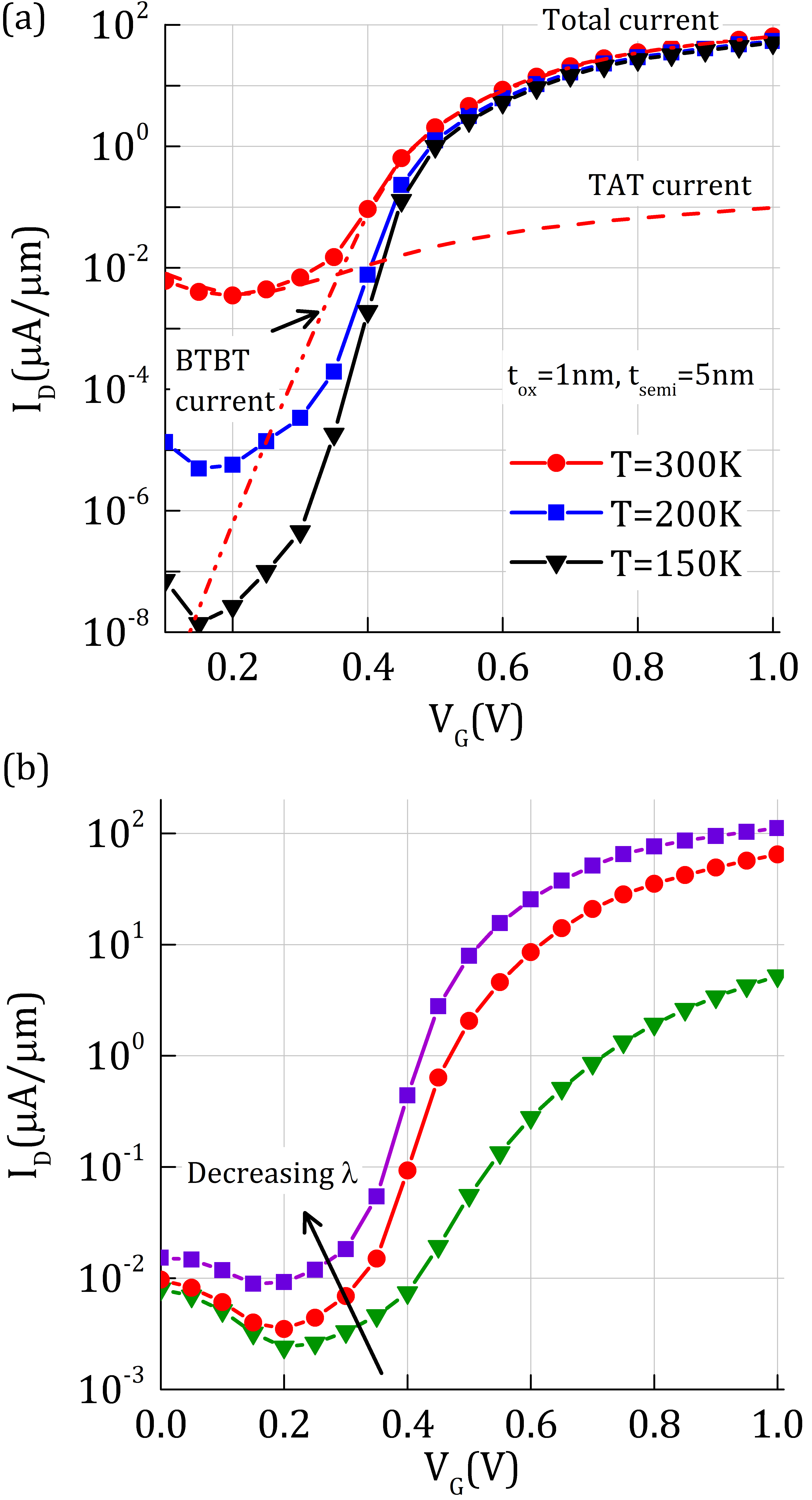}
\caption{(a) Total (TAT+BTBT) current in In$_{0.53}$Ga$_{0.47}$As based homojunction TFET with the device structure as shown in Fig. 1 with EOT, $t_\mathrm{ox}$ and semiconductor thickness $t_\mathrm{semi}$ 1 nm and 5 nm respectively and a drain bias of V$_\mathrm{DS}$ = 0.3 V. BTBT follows WKB formalism above threshold (when the bands overlap), while below threshold the BTBT has an exponentially decaying transmission due to the band tails (Urbach tails, in this case at 40 mV/dec at 300 K, 25 mV/dec at 150 K). TAT is temperature dependent and obscures the steepest part of the BTBT current in the subthreshold regime ($\sim V_G<0.4$ V) for temperatures above 150 K. Midgap $D_\mathrm{it}$ is assumed to be 5x10$^{12}$/cm$^2$-eV. (b) Since both TAT and BTBT are electric field dependent, the thickness of the oxide and the semiconductor affects the current levels as well as the subthreshold slope. In this calculation, gradually decrease (from bottom to top) the thicknesses resulting in decreasing scaling lengths $\lambda$. $t_\mathrm{ox}$ is 2 nm, 1 nm and 0.75 nm and $t_\mathrm{semi}$ is 10 nm, 5 nm and 1 nm respectively. Even for very thin oxide and body thickness, TAT is large enough to overshadow the steep change of BTBT.}
\label{fig:lambda}
\end{figure}
We apply the model for the top gate structure shown in Fig. \ref{fig:device}a. Effective oxide thickness (EOT), $t_\mathrm{ox}$ and semiconductor body thickness $t_\mathrm{semi}$ are 1 nm and 5 nm respectively. $D_\mathrm{it}$ profile in Ref. \cite{brammertz2009} is used for III-V with midgap $D_\mathrm{it}$ of $5\times10^{12}/$cm$^{2}$-eV. Although $D_\mathrm{it}$ is a function of energy in the bandgap, we found that in most cases the midgap trap density dominates the trap current. Channel length, $L_\mathrm{ch}$ is 100 nm. Source and drain contact regions are degenerately doped while the channel is undoped. The capture cross-section for electrons, holes are $\sigma_n = 5\times 10^{-17}\,\mathrm{m}^2$, $\sigma_p = 5\times 10^{-18}\,\mathrm{m}^2$  \cite{furlan2001,selberherr2012} and the carrier lifetimes are calculated from there using the thermal velocity, $v_{th} = \sqrt{\frac{8KT}{\pi m^*}}$. An underlap (10 nm long) at the channel-drain end is used to suppress the electric field in the drain end and therefore the ambipolarity. For the transfer curves, we use a drain bias $V_{DS} = 0.3$ V. We ignore channel resistance due to carrier scattering in the channel since the resistance due to TAT and BTBT is substantially higher. 

Fig. 3a shows the transfer plots for In$_{0.53}$Ga$_{0.47}$As TFET at various temperatures. For room temperature, the TAT and BTBT current components of the total current are also shown. Well above the threshold voltage ($V_t \sim 0.37$ V), the total current mainly comes from BTBT. The TAT current is just enough so that it intersects with the BTBT current near the threshold voltage, therefore the total current below $V_t$ is dominated by the TAT.  The TAT thus obscures the steepest part of the BTBT ($\sim$ 40 mV/dec in this calculation) and so the minimum subthreshold swing ($\sim$ 75 mV/dec) is limited by the rate of change of BTBT current just above the threshold voltage. This subthreshold swing will get much worse for thicker oxide and body thickness. Such transfer behavior with a valley near the minimum current is seen in most experiments on III-V TFETs \cite{tao2015,mohata2012,pandey2015}. At lower temperatures, electron hole generation rate is reduced leading to lower TAT. For temperatures lower than 200 K, intrinsic subthreshold swing is observed. The current above the threshold voltage is weakly dependent on temperature while current below the threshold varies stongly with temperature.  
In other words, the lowest achievable current at any given temperature is a function of temperature (decreases from $\sim$ 1 nA at 300 K to $\sim$ 10 fA at 150 K), similar to what is seen in the experiments \cite{mookerjea2010,tao2015,zhao2014,mookerjea2009,noguchi2013}.

To demonstrate the effect of the scaling length $\lambda$ and the local electric field, Fig. 3b shows the transfer plots for different oxide and body thicknesses at T = 300 K. With $t_\mathrm{ox}=0.75$ nm and $t_\mathrm{semi}=1$ nm (violet squares), ON current increases substantially due to the increase of the local electric field near the source.  Subthreshold swing also improves  to $\sim$ 65 mV/dec which is still not subthermal. This is due to the fact that the TAT current has also increased, thus limiting the advantage of the higher electric field. We infer that the same effect takes place in heterojunction TFETs, making it difficult to observe subthermal switching for those structures as well. 

\begin{figure}
\centering
\includegraphics[width=3.4in]{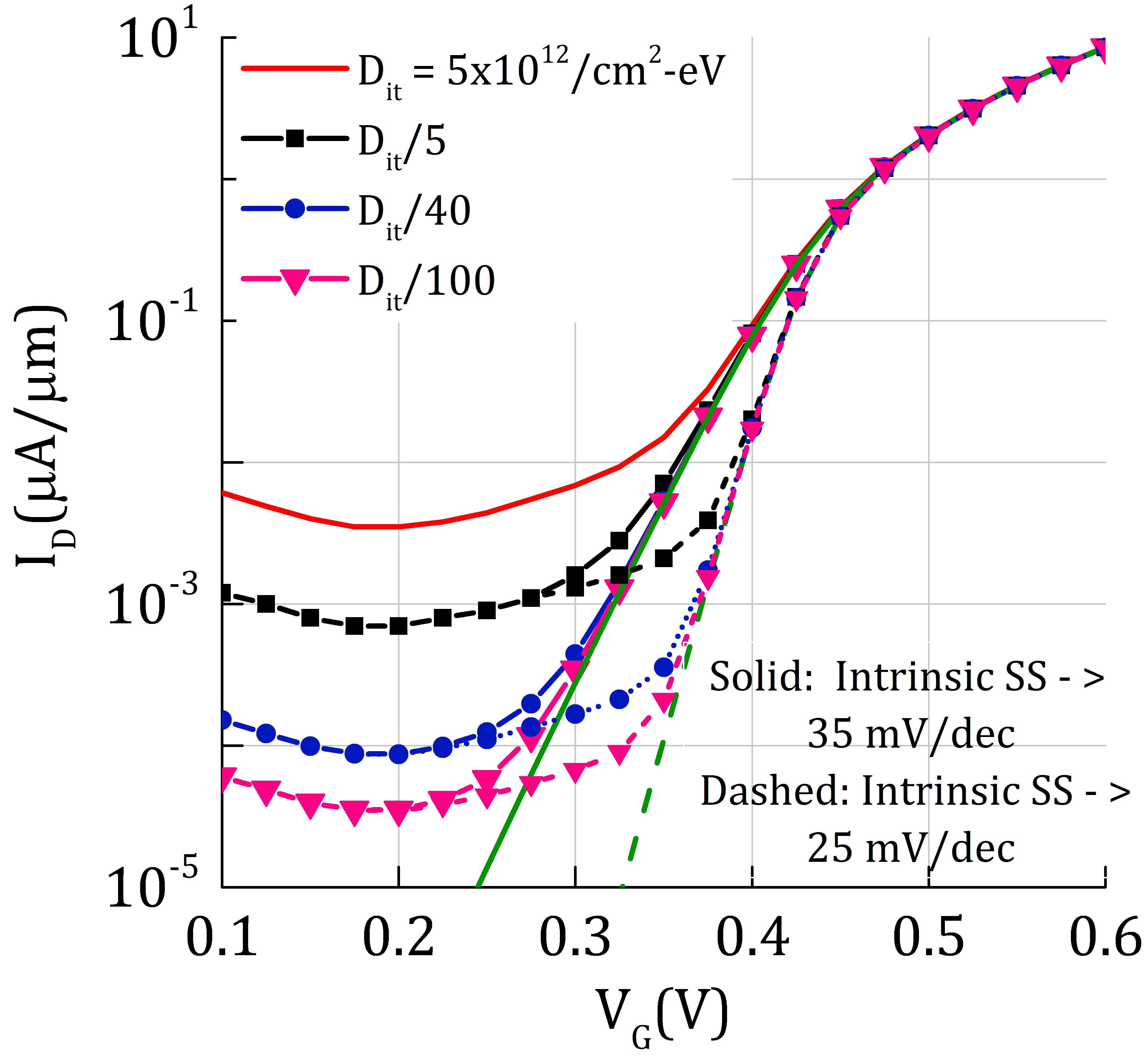}
\caption{Impact of $D_\mathrm{it}$ magnitude on the transfer characteristics for $t_\mathrm{ox}$ = 1 nm, $t_\mathrm{semi}$ = 5 nm for two different Urbach tail parameter $E_0$. Total current at different midgap $D_\mathrm{it}$ levels. At roughly 10$^{11}$/cm$^2$-eV (typical $D_\mathrm{it}$/40), the TAT current is low enough for the steep BTBT current to be manifested (for two orders of magnitude at $\sim$ 40 mV/dec and $\sim$ 28 mV/dec). 
}
\label{fig:dit}
\end{figure}
Fig. 4 shows the transfer plots for various trap density for two different intrinsic subthreshold swings (Urbach tails at 35 and 25 mV/dec) with a motivation to find the trap density required to achieve subthermal switching for multiple decades. We find that a trap density of 1.25$\times$10$^{11}/\mathrm{cm}^{2}$-eV, which is about 40 times smaller than today's typical midgap trap density, achieves about two orders of current change at $\sim$ 40 mV/dec. For the steeper intrinsic swing, we again get two orders of current change at subthermal rate ($\sim$ 28 mV/dec). In this case, the TAT and BTBT intersects at a higher $V_G$. Since the TAT increases with $V_G$, a steeper Urbach tail does not necessarily increase the ON-OFF ratio (at subthermal rate). Therefore the ON-OFF ratio at subthermal rate is determined mainly by the trap density, while the subthreshold swing is determined by the Urbach tails. We see this again when the trap density is reduced by 100 times, where we get about three orders of change in current at subthermal rate for both Urbach tails. 

We applied the same model to silicon to see how the transfer characteristic changes at reduced trap density and different material properties. In Fig. 5, we see that both BTBT and TAT decrease substantially due to heavier effective mass and higher bandgap with a midgap trap density of $5\times 10^{10}$/cm$^2$-eV, which is typical in today's silicon technology. Similar to III-V, the steepest part of the BTBT is not seen due to the TAT. However at $1\times 10^{10}$/cm$^2$-eV, we found (dashed line) two orders of current change at 50 mV/dec (in pA range). Such $D_\mathrm{it}$ is much easier to achieve in silicon than the requirements mentioned earlier for III-V. This also explains, why most experiments reporting margin subthermal switching at very low currents involved silicon, where it is likely that such trap density may be achieved. 
\begin{figure}
\centering
\includegraphics[width=3.4in]{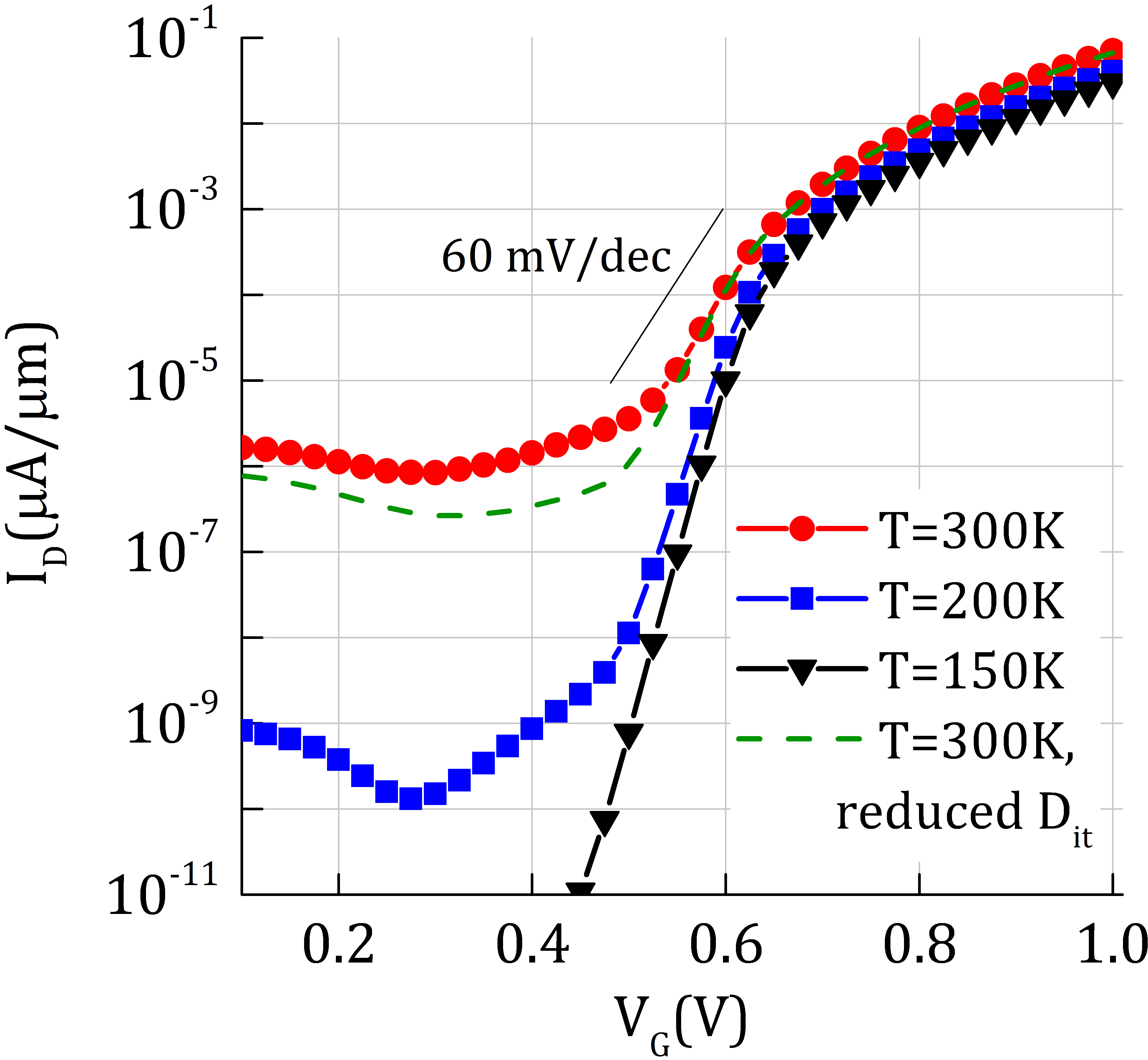}
\caption{
TAT+BTBT current at different temperatures for homojunction silicon TFET $t_\mathrm{ox}$ = 1 nm, $t_\mathrm{semi}$ = 5 nm.  Since silicon has much lower $D_\mathrm{it}$ (mid gap density assumed here is 5x10$^{10}$/cm$^2$-eV), TAT to BTBT transition takes place at a higher temperature ($\sim$ 200 K) compared to III-V. For a slightly lower midgap $D_\mathrm{it}$ of 1$\times10^{11}/\mathrm{cm}^2$-eV (dashed), we get two orders of current change at 50 mV/dec at room temperature. }
\label{fig:ss}
\end{figure}
\section{Conclusion}
We provide an analysis of the parasitic trap assisted tunneling current in TFETs. We show that in most cases, the subthreshold current in TFETs is dominated by TAT, regardless of channel material. The takeover from TAT to band to band tunneling depends on the temperature, electrostatic characteristic length, material parameters (e.g. effective mass) and the rate of change of the exponential band tails (Urbach tails). We show that engineering efforts to increase the ON current are also likely to increase the subthreshold current, since both BTBT and TAT are driven by the same mechanism (tunneling through a barrier). The TAT current is much more deleterious than just the electron-phonon scattering without traps. We find that to get a reasonable ON-OFF ratio with steeper than 60 mV/dec subthreshold swing at room temperature, trap density has to be reduced by 40-100 times  compared to the state of the art for III-V semiconductors, for reasonable structural device parameters.  

\section{Acknowledgment}

This work was supported by National Science Foundation under the Center for Energy Efficient Electronics Science Center, Award 0939514, and the NCN-NEEDS Program, Grant 1227020-EEC, with additional support by the Semiconductor Research Corporation. Authors also thank Eli Yablonovitch (UC Berkeley), Patrick Xiao (UC Berkeley), Sapan Agarwal (Sandia), Ujwal Radhakrishna (MIT), Alan Seabaugh (University of Notre Dame) for useful discussions.

\bibliographystyle{apsrev4-1}
\bibliography{39bib}

\begin{thebibliography}{46}
\expandafter\ifx\csname natexlab\endcsname\relax\def\natexlab#1{#1}\fi
\expandafter\ifx\csname bibnamefont\endcsname\relax
  \def\bibnamefont#1{#1}\fi
\expandafter\ifx\csname bibfnamefont\endcsname\relax
  \def\bibfnamefont#1{#1}\fi
\expandafter\ifx\csname citenamefont\endcsname\relax
  \def\citenamefont#1{#1}\fi
\expandafter\ifx\csname url\endcsname\relax
  \def\url#1{\texttt{#1}}\fi
\expandafter\ifx\csname urlprefix\endcsname\relax\def\urlprefix{URL }\fi
\providecommand{\bibinfo}[2]{#2}
\providecommand{\eprint}[2][]{\url{#2}}

\bibitem[{\citenamefont{Seabaugh and Zhang}(2010)}]{seabaugh2010}
\bibinfo{author}{\bibfnamefont{A.~C.} \bibnamefont{Seabaugh}} \bibnamefont{and}
  \bibinfo{author}{\bibfnamefont{Q.}~\bibnamefont{Zhang}},
  \bibinfo{journal}{Proceedings of the IEEE} \textbf{\bibinfo{volume}{98}},
  \bibinfo{pages}{2095} (\bibinfo{year}{2010}).

\bibitem[{\citenamefont{Avci et~al.}(2013)\citenamefont{Avci, Morris, Hasan,
  Kotlyar, Kim, Rios, Nikonov, Young et~al.}}]{avci2013}
\bibinfo{author}{\bibfnamefont{U.~E.} \bibnamefont{Avci}},
  \bibinfo{author}{\bibfnamefont{D.~H.} \bibnamefont{Morris}},
  \bibinfo{author}{\bibfnamefont{S.}~\bibnamefont{Hasan}},
  \bibinfo{author}{\bibfnamefont{R.}~\bibnamefont{Kotlyar}},
  \bibinfo{author}{\bibfnamefont{R.}~\bibnamefont{Kim}},
  \bibinfo{author}{\bibfnamefont{R.}~\bibnamefont{Rios}},
  \bibinfo{author}{\bibfnamefont{D.~E.} \bibnamefont{Nikonov}},
  \bibinfo{author}{\bibfnamefont{I.}~\bibnamefont{Young}},
  \bibnamefont{et~al.}, in \emph{\bibinfo{booktitle}{Electron Devices Meeting
  (IEDM), 2013 IEEE International}} (\bibinfo{organization}{IEEE},
  \bibinfo{year}{2013}), pp. \bibinfo{pages}{33--4}.

\bibitem[{\citenamefont{Young et~al.}(2015)\citenamefont{Young, Avci, and
  Morris}}]{young2015}
\bibinfo{author}{\bibfnamefont{I.}~\bibnamefont{Young}},
  \bibinfo{author}{\bibfnamefont{U.}~\bibnamefont{Avci}}, \bibnamefont{and}
  \bibinfo{author}{\bibfnamefont{D.}~\bibnamefont{Morris}}, in
  \emph{\bibinfo{booktitle}{Electron Devices Meeting (IEDM), 2015 IEEE
  International}} (\bibinfo{organization}{IEEE}, \bibinfo{year}{2015}), pp.
  \bibinfo{pages}{25--5}.

\bibitem[{\citenamefont{Knoch and Appenzeller}(2010)}]{knoch2010}
\bibinfo{author}{\bibfnamefont{J.}~\bibnamefont{Knoch}} \bibnamefont{and}
  \bibinfo{author}{\bibfnamefont{J.}~\bibnamefont{Appenzeller}},
  \bibinfo{journal}{Electron Device Letters, IEEE}
  \textbf{\bibinfo{volume}{31}}, \bibinfo{pages}{305} (\bibinfo{year}{2010}).

\bibitem[{\citenamefont{Appenzeller et~al.}(2004)\citenamefont{Appenzeller,
  Lin, Knoch, and Avouris}}]{appenzeller_04}
\bibinfo{author}{\bibfnamefont{J.}~\bibnamefont{Appenzeller}},
  \bibinfo{author}{\bibfnamefont{Y.-M.} \bibnamefont{Lin}},
  \bibinfo{author}{\bibfnamefont{J.}~\bibnamefont{Knoch}}, \bibnamefont{and}
  \bibinfo{author}{\bibfnamefont{P.}~\bibnamefont{Avouris}},
  \bibinfo{journal}{Phys. Rev. Lett.} \textbf{\bibinfo{volume}{93}},
  \bibinfo{pages}{196805} (\bibinfo{year}{2004}).

\bibitem[{\citenamefont{Choi et~al.}(2007)\citenamefont{Choi, Park, Lee, and
  Liu}}]{choi2007}
\bibinfo{author}{\bibfnamefont{W.~Y.} \bibnamefont{Choi}},
  \bibinfo{author}{\bibfnamefont{B.-G.} \bibnamefont{Park}},
  \bibinfo{author}{\bibfnamefont{J.~D.} \bibnamefont{Lee}}, \bibnamefont{and}
  \bibinfo{author}{\bibfnamefont{T.-J.~K.} \bibnamefont{Liu}},
  \bibinfo{journal}{Electron Device Letters, IEEE}
  \textbf{\bibinfo{volume}{28}}, \bibinfo{pages}{743} (\bibinfo{year}{2007}).

\bibitem[{\citenamefont{Krishnamohan et~al.}(2008)\citenamefont{Krishnamohan,
  Kim, Raghunathan, and Saraswat}}]{krishnamohan2008}
\bibinfo{author}{\bibfnamefont{T.}~\bibnamefont{Krishnamohan}},
  \bibinfo{author}{\bibfnamefont{D.}~\bibnamefont{Kim}},
  \bibinfo{author}{\bibfnamefont{S.}~\bibnamefont{Raghunathan}},
  \bibnamefont{and} \bibinfo{author}{\bibfnamefont{K.}~\bibnamefont{Saraswat}},
  in \emph{\bibinfo{booktitle}{Electron Devices Meeting, 2008. IEDM 2008. IEEE
  International}} (\bibinfo{organization}{IEEE}, \bibinfo{year}{2008}), pp.
  \bibinfo{pages}{1--3}.

\bibitem[{\citenamefont{Jeon et~al.}(2010)\citenamefont{Jeon, Loh, Patel, Kang,
  Oh, Bowonder, Park, Park, Smith, Majhi et~al.}}]{jeon2010}
\bibinfo{author}{\bibfnamefont{K.}~\bibnamefont{Jeon}},
  \bibinfo{author}{\bibfnamefont{W.-Y.} \bibnamefont{Loh}},
  \bibinfo{author}{\bibfnamefont{P.}~\bibnamefont{Patel}},
  \bibinfo{author}{\bibfnamefont{C.~Y.} \bibnamefont{Kang}},
  \bibinfo{author}{\bibfnamefont{J.}~\bibnamefont{Oh}},
  \bibinfo{author}{\bibfnamefont{A.}~\bibnamefont{Bowonder}},
  \bibinfo{author}{\bibfnamefont{C.}~\bibnamefont{Park}},
  \bibinfo{author}{\bibfnamefont{C.}~\bibnamefont{Park}},
  \bibinfo{author}{\bibfnamefont{C.}~\bibnamefont{Smith}},
  \bibinfo{author}{\bibfnamefont{P.}~\bibnamefont{Majhi}},
  \bibnamefont{et~al.}, in \emph{\bibinfo{booktitle}{VLSI technology (VLSIT),
  2010 symposium on}} (\bibinfo{organization}{IEEE}, \bibinfo{year}{2010}), pp.
  \bibinfo{pages}{121--122}.

\bibitem[{\citenamefont{Dewey et~al.}(2011)\citenamefont{Dewey, Chu-Kung,
  Boardman, Fastenau, Kavalieros, Kotlyar, Liu, Lubyshev, Metz, Mukherjee
  et~al.}}]{dewey2011}
\bibinfo{author}{\bibfnamefont{G.}~\bibnamefont{Dewey}},
  \bibinfo{author}{\bibfnamefont{B.}~\bibnamefont{Chu-Kung}},
  \bibinfo{author}{\bibfnamefont{J.}~\bibnamefont{Boardman}},
  \bibinfo{author}{\bibfnamefont{J.}~\bibnamefont{Fastenau}},
  \bibinfo{author}{\bibfnamefont{J.}~\bibnamefont{Kavalieros}},
  \bibinfo{author}{\bibfnamefont{R.}~\bibnamefont{Kotlyar}},
  \bibinfo{author}{\bibfnamefont{W.}~\bibnamefont{Liu}},
  \bibinfo{author}{\bibfnamefont{D.}~\bibnamefont{Lubyshev}},
  \bibinfo{author}{\bibfnamefont{M.}~\bibnamefont{Metz}},
  \bibinfo{author}{\bibfnamefont{N.}~\bibnamefont{Mukherjee}},
  \bibnamefont{et~al.}, in \emph{\bibinfo{booktitle}{Electron Devices Meeting
  (IEDM), 2011 IEEE International}} (\bibinfo{organization}{IEEE},
  \bibinfo{year}{2011}), pp. \bibinfo{pages}{33--6}.

\bibitem[{\citenamefont{Gandhi et~al.}(2011)\citenamefont{Gandhi, Chen, Singh,
  Banerjee, and Lee}}]{gandhi2011}
\bibinfo{author}{\bibfnamefont{R.}~\bibnamefont{Gandhi}},
  \bibinfo{author}{\bibfnamefont{Z.}~\bibnamefont{Chen}},
  \bibinfo{author}{\bibfnamefont{N.}~\bibnamefont{Singh}},
  \bibinfo{author}{\bibfnamefont{K.}~\bibnamefont{Banerjee}}, \bibnamefont{and}
  \bibinfo{author}{\bibfnamefont{S.}~\bibnamefont{Lee}},
  \bibinfo{journal}{Electron Device Letters, IEEE}
  \textbf{\bibinfo{volume}{32}}, \bibinfo{pages}{437} (\bibinfo{year}{2011}).

\bibitem[{\citenamefont{Ganjipour et~al.}(2012)\citenamefont{Ganjipour,
  Wallentin, Borgström, Samuelson, and Thelander}}]{ganjipour2012}
\bibinfo{author}{\bibfnamefont{B.}~\bibnamefont{Ganjipour}},
  \bibinfo{author}{\bibfnamefont{J.}~\bibnamefont{Wallentin}},
  \bibinfo{author}{\bibfnamefont{M.~T.} \bibnamefont{Borgström}},
  \bibinfo{author}{\bibfnamefont{L.}~\bibnamefont{Samuelson}},
  \bibnamefont{and}
  \bibinfo{author}{\bibfnamefont{C.}~\bibnamefont{Thelander}},
  \bibinfo{journal}{ACS nano} \textbf{\bibinfo{volume}{6}},
  \bibinfo{pages}{3109} (\bibinfo{year}{2012}).

\bibitem[{\citenamefont{Mookerjea et~al.}(2010)\citenamefont{Mookerjea, Mohata,
  Mayer, Narayanan, and Datta}}]{mookerjea2010}
\bibinfo{author}{\bibfnamefont{S.}~\bibnamefont{Mookerjea}},
  \bibinfo{author}{\bibfnamefont{D.}~\bibnamefont{Mohata}},
  \bibinfo{author}{\bibfnamefont{T.}~\bibnamefont{Mayer}},
  \bibinfo{author}{\bibfnamefont{V.}~\bibnamefont{Narayanan}},
  \bibnamefont{and} \bibinfo{author}{\bibfnamefont{S.}~\bibnamefont{Datta}},
  \bibinfo{journal}{IEEE Electron Device Letters}
  \textbf{\bibinfo{volume}{31}}, \bibinfo{pages}{564} (\bibinfo{year}{2010}).

\bibitem[{\citenamefont{Yu et~al.}(2015)\citenamefont{Yu, Radhakrishna, Hoyt,
  and Antoniadis}}]{tao2015}
\bibinfo{author}{\bibfnamefont{T.}~\bibnamefont{Yu}},
  \bibinfo{author}{\bibfnamefont{U.}~\bibnamefont{Radhakrishna}},
  \bibinfo{author}{\bibfnamefont{J.~L.} \bibnamefont{Hoyt}}, \bibnamefont{and}
  \bibinfo{author}{\bibfnamefont{D.~A.} \bibnamefont{Antoniadis}}, in
  \emph{\bibinfo{booktitle}{Electron Devices Meeting (IEDM), 2014 IEEE
  International}} (\bibinfo{organization}{IEEE}, \bibinfo{year}{2015}).

\bibitem[{\citenamefont{Zhao et~al.}(2014)\citenamefont{Zhao, Vardi, del Alamo
  et~al.}}]{zhao2014}
\bibinfo{author}{\bibfnamefont{X.}~\bibnamefont{Zhao}},
  \bibinfo{author}{\bibfnamefont{A.}~\bibnamefont{Vardi}},
  \bibinfo{author}{\bibfnamefont{J.}~\bibnamefont{del Alamo}},
  \bibnamefont{et~al.}, in \emph{\bibinfo{booktitle}{Electron Devices Meeting
  (IEDM), 2014 IEEE International}} (\bibinfo{organization}{IEEE},
  \bibinfo{year}{2014}), pp. \bibinfo{pages}{25--5}.

\bibitem[{\citenamefont{Sarkar et~al.}(2015)\citenamefont{Sarkar, Xie, Liu,
  Cao, Kang, Gong, Kraemer, Ajayan, and Banerjee}}]{sarkar2015}
\bibinfo{author}{\bibfnamefont{D.}~\bibnamefont{Sarkar}},
  \bibinfo{author}{\bibfnamefont{X.}~\bibnamefont{Xie}},
  \bibinfo{author}{\bibfnamefont{W.}~\bibnamefont{Liu}},
  \bibinfo{author}{\bibfnamefont{W.}~\bibnamefont{Cao}},
  \bibinfo{author}{\bibfnamefont{J.}~\bibnamefont{Kang}},
  \bibinfo{author}{\bibfnamefont{Y.}~\bibnamefont{Gong}},
  \bibinfo{author}{\bibfnamefont{S.}~\bibnamefont{Kraemer}},
  \bibinfo{author}{\bibfnamefont{P.~M.} \bibnamefont{Ajayan}},
  \bibnamefont{and} \bibinfo{author}{\bibfnamefont{K.}~\bibnamefont{Banerjee}},
  \bibinfo{journal}{Nature} \textbf{\bibinfo{volume}{526}}, \bibinfo{pages}{91}
  (\bibinfo{year}{2015}).

\bibitem[{\citenamefont{Roy et~al.}(2015)\citenamefont{Roy, Tosun, Cao, Fang,
  Lien, Zhao, Chen, Chueh, Guo, and Javey}}]{roy2015}
\bibinfo{author}{\bibfnamefont{T.}~\bibnamefont{Roy}},
  \bibinfo{author}{\bibfnamefont{M.}~\bibnamefont{Tosun}},
  \bibinfo{author}{\bibfnamefont{X.}~\bibnamefont{Cao}},
  \bibinfo{author}{\bibfnamefont{H.}~\bibnamefont{Fang}},
  \bibinfo{author}{\bibfnamefont{D.-H.} \bibnamefont{Lien}},
  \bibinfo{author}{\bibfnamefont{P.}~\bibnamefont{Zhao}},
  \bibinfo{author}{\bibfnamefont{Y.-Z.} \bibnamefont{Chen}},
  \bibinfo{author}{\bibfnamefont{Y.-L.} \bibnamefont{Chueh}},
  \bibinfo{author}{\bibfnamefont{J.}~\bibnamefont{Guo}}, \bibnamefont{and}
  \bibinfo{author}{\bibfnamefont{A.}~\bibnamefont{Javey}},
  \bibinfo{journal}{ACS nano} \textbf{\bibinfo{volume}{9}},
  \bibinfo{pages}{2071} (\bibinfo{year}{2015}).

\bibitem[{\citenamefont{Nourbakhsh et~al.}(2016)\citenamefont{Nourbakhsh,
  Zubair, Dresselhaus, and Palacios}}]{amir2016}
\bibinfo{author}{\bibfnamefont{A.}~\bibnamefont{Nourbakhsh}},
  \bibinfo{author}{\bibfnamefont{A.}~\bibnamefont{Zubair}},
  \bibinfo{author}{\bibfnamefont{M.~S.} \bibnamefont{Dresselhaus}},
  \bibnamefont{and} \bibinfo{author}{\bibfnamefont{T.}~\bibnamefont{Palacios}},
  \bibinfo{journal}{Nano letters}  (\bibinfo{year}{2016}).

\bibitem[{\citenamefont{Schenk}(1992)}]{schenk1992}
\bibinfo{author}{\bibfnamefont{A.}~\bibnamefont{Schenk}},
  \bibinfo{journal}{Solid-State Electronics} \textbf{\bibinfo{volume}{35}},
  \bibinfo{pages}{1585} (\bibinfo{year}{1992}).

\bibitem[{\citenamefont{Hurkx et~al.}(1992)\citenamefont{Hurkx, Klaassen, and
  Knuvers}}]{hurkx1992}
\bibinfo{author}{\bibfnamefont{G.}~\bibnamefont{Hurkx}},
  \bibinfo{author}{\bibfnamefont{D.}~\bibnamefont{Klaassen}}, \bibnamefont{and}
  \bibinfo{author}{\bibfnamefont{M.}~\bibnamefont{Knuvers}},
  \bibinfo{journal}{Electron Devices, IEEE Transactions on}
  \textbf{\bibinfo{volume}{39}}, \bibinfo{pages}{331} (\bibinfo{year}{1992}).

\bibitem[{\citenamefont{Khayer and Lake}(2011)}]{khayer2011}
\bibinfo{author}{\bibfnamefont{M.~A.} \bibnamefont{Khayer}} \bibnamefont{and}
  \bibinfo{author}{\bibfnamefont{R.~K.} \bibnamefont{Lake}},
  \bibinfo{journal}{Journal of Applied Physics} \textbf{\bibinfo{volume}{110}},
  \bibinfo{pages}{074508} (\bibinfo{year}{2011}).

\bibitem[{\citenamefont{Koswatta et~al.}(2008)\citenamefont{Koswatta,
  Lundstrom, and Nikonov}}]{koswatta2008}
\bibinfo{author}{\bibfnamefont{S.~O.} \bibnamefont{Koswatta}},
  \bibinfo{author}{\bibfnamefont{M.~S.} \bibnamefont{Lundstrom}},
  \bibnamefont{and} \bibinfo{author}{\bibfnamefont{D.~E.}
  \bibnamefont{Nikonov}}, \bibinfo{journal}{Applied Physics Letters}
  \textbf{\bibinfo{volume}{92}}, \bibinfo{pages}{043125}
  (\bibinfo{year}{2008}).

\bibitem[{\citenamefont{Vallett et~al.}(2010)\citenamefont{Vallett, Minassian,
  Kaszuba, Datta, Redwing, and Mayer}}]{vallett2010}
\bibinfo{author}{\bibfnamefont{A.~L.} \bibnamefont{Vallett}},
  \bibinfo{author}{\bibfnamefont{S.}~\bibnamefont{Minassian}},
  \bibinfo{author}{\bibfnamefont{P.}~\bibnamefont{Kaszuba}},
  \bibinfo{author}{\bibfnamefont{S.}~\bibnamefont{Datta}},
  \bibinfo{author}{\bibfnamefont{J.~M.} \bibnamefont{Redwing}},
  \bibnamefont{and} \bibinfo{author}{\bibfnamefont{T.~S.} \bibnamefont{Mayer}},
  \bibinfo{journal}{Nano letters} \textbf{\bibinfo{volume}{10}},
  \bibinfo{pages}{4813} (\bibinfo{year}{2010}).

\bibitem[{\citenamefont{Pala and Esseni}(2013)}]{pala2013}
\bibinfo{author}{\bibfnamefont{M.~G.} \bibnamefont{Pala}} \bibnamefont{and}
  \bibinfo{author}{\bibfnamefont{D.}~\bibnamefont{Esseni}},
  \bibinfo{journal}{Electron Devices, IEEE Transactions on}
  \textbf{\bibinfo{volume}{60}}, \bibinfo{pages}{2795} (\bibinfo{year}{2013}).

\bibitem[{\citenamefont{Qiu et~al.}(2014)\citenamefont{Qiu, Wang, Huang, and
  Huang}}]{qiu2014}
\bibinfo{author}{\bibfnamefont{Y.}~\bibnamefont{Qiu}},
  \bibinfo{author}{\bibfnamefont{R.}~\bibnamefont{Wang}},
  \bibinfo{author}{\bibfnamefont{Q.}~\bibnamefont{Huang}}, \bibnamefont{and}
  \bibinfo{author}{\bibfnamefont{R.}~\bibnamefont{Huang}},
  \bibinfo{journal}{Electron Devices, IEEE Transactions on}
  \textbf{\bibinfo{volume}{61}}, \bibinfo{pages}{1284} (\bibinfo{year}{2014}).

\bibitem[{\citenamefont{Avci et~al.}(2015)\citenamefont{Avci, Chu-kung,
  Agrawal, Dewey, Le, Rios, Morris, Hasan, Kotlyar, Kavalieros
  et~al.}}]{avci2015}
\bibinfo{author}{\bibfnamefont{U.}~\bibnamefont{Avci}},
  \bibinfo{author}{\bibfnamefont{B.}~\bibnamefont{Chu-kung}},
  \bibinfo{author}{\bibfnamefont{A.}~\bibnamefont{Agrawal}},
  \bibinfo{author}{\bibfnamefont{G.}~\bibnamefont{Dewey}},
  \bibinfo{author}{\bibfnamefont{V.}~\bibnamefont{Le}},
  \bibinfo{author}{\bibfnamefont{R.}~\bibnamefont{Rios}},
  \bibinfo{author}{\bibfnamefont{D.}~\bibnamefont{Morris}},
  \bibinfo{author}{\bibfnamefont{S.}~\bibnamefont{Hasan}},
  \bibinfo{author}{\bibfnamefont{R.}~\bibnamefont{Kotlyar}},
  \bibinfo{author}{\bibfnamefont{J.}~\bibnamefont{Kavalieros}},
  \bibnamefont{et~al.}, in \emph{\bibinfo{booktitle}{Electron Devices Meeting
  (IEDM), 2015 IEEE International}} (\bibinfo{organization}{IEEE},
  \bibinfo{year}{2015}), pp. \bibinfo{pages}{25--5}.

\bibitem[{\citenamefont{Agarwal and Yablonovitch}(2015)}]{sapan2015}
\bibinfo{author}{\bibfnamefont{S.}~\bibnamefont{Agarwal}} \bibnamefont{and}
  \bibinfo{author}{\bibfnamefont{E.}~\bibnamefont{Yablonovitch}}, in
  \emph{\bibinfo{booktitle}{Device Research Conference (DRC), 2015 73rd
  Annual}} (\bibinfo{organization}{IEEE}, \bibinfo{year}{2015}), pp.
  \bibinfo{pages}{247--248}.

\bibitem[{\citenamefont{Furlan}(2001)}]{furlan2001}
\bibinfo{author}{\bibfnamefont{J.}~\bibnamefont{Furlan}},
  \bibinfo{journal}{Progress in quantum electronics}
  \textbf{\bibinfo{volume}{25}}, \bibinfo{pages}{55} (\bibinfo{year}{2001}).

\bibitem[{\citenamefont{Sze and Ng}(2006)}]{sze2006}
\bibinfo{author}{\bibfnamefont{S.~M.} \bibnamefont{Sze}} \bibnamefont{and}
  \bibinfo{author}{\bibfnamefont{K.~K.} \bibnamefont{Ng}},
  \emph{\bibinfo{title}{Physics of semiconductor devices}}
  (\bibinfo{publisher}{John Wiley \& Sons}, \bibinfo{year}{2006}).

\bibitem[{\citenamefont{Woo et~al.}(1987)\citenamefont{Woo, Plummer, and
  Stork}}]{woo1987}
\bibinfo{author}{\bibfnamefont{J.}~\bibnamefont{Woo}},
  \bibinfo{author}{\bibfnamefont{J.~D.} \bibnamefont{Plummer}},
  \bibnamefont{and} \bibinfo{author}{\bibfnamefont{J.}~\bibnamefont{Stork}},
  \bibinfo{journal}{Electron Devices, IEEE Transactions on}
  \textbf{\bibinfo{volume}{34}}, \bibinfo{pages}{130} (\bibinfo{year}{1987}).

\bibitem[{\citenamefont{Pelaz et~al.}(1994)\citenamefont{Pelaz, Orantes,
  Vincente, Bailon, and Barbolla}}]{pelaz1994}
\bibinfo{author}{\bibfnamefont{L.}~\bibnamefont{Pelaz}},
  \bibinfo{author}{\bibfnamefont{J.~L.} \bibnamefont{Orantes}},
  \bibinfo{author}{\bibfnamefont{J.}~\bibnamefont{Vincente}},
  \bibinfo{author}{\bibfnamefont{L.~A.} \bibnamefont{Bailon}},
  \bibnamefont{and} \bibinfo{author}{\bibfnamefont{J.}~\bibnamefont{Barbolla}},
  \bibinfo{journal}{IEEE Transactions on Electron Devices}
  \textbf{\bibinfo{volume}{41}}, \bibinfo{pages}{587} (\bibinfo{year}{1994}).

\bibitem[{\citenamefont{Huang et~al.}(1997)\citenamefont{Huang, Qin, Zhang,
  Sin, and Poon}}]{huang1997}
\bibinfo{author}{\bibfnamefont{Q.-A.} \bibnamefont{Huang}},
  \bibinfo{author}{\bibfnamefont{M.}~\bibnamefont{Qin}},
  \bibinfo{author}{\bibfnamefont{B.}~\bibnamefont{Zhang}},
  \bibinfo{author}{\bibfnamefont{J.~K.} \bibnamefont{Sin}}, \bibnamefont{and}
  \bibinfo{author}{\bibfnamefont{M.}~\bibnamefont{Poon}},
  \bibinfo{journal}{IEEE electron device letters}
  \textbf{\bibinfo{volume}{18}}, \bibinfo{pages}{616} (\bibinfo{year}{1997}).

\bibitem[{\citenamefont{Yoon and Salahuddin}(2012)}]{yoon2012}
\bibinfo{author}{\bibfnamefont{Y.}~\bibnamefont{Yoon}} \bibnamefont{and}
  \bibinfo{author}{\bibfnamefont{S.}~\bibnamefont{Salahuddin}},
  \bibinfo{journal}{Applied Physics Letters} \textbf{\bibinfo{volume}{101}},
  \bibinfo{pages}{263501} (\bibinfo{year}{2012}).

\bibitem[{\citenamefont{Yan et~al.}(1992)\citenamefont{Yan, Ourmazd, and
  Lee}}]{yan1992}
\bibinfo{author}{\bibfnamefont{R.-H.} \bibnamefont{Yan}},
  \bibinfo{author}{\bibfnamefont{A.}~\bibnamefont{Ourmazd}}, \bibnamefont{and}
  \bibinfo{author}{\bibfnamefont{K.~F.} \bibnamefont{Lee}},
  \bibinfo{journal}{Electron Devices, IEEE Transactions on}
  \textbf{\bibinfo{volume}{39}}, \bibinfo{pages}{1704} (\bibinfo{year}{1992}).

\bibitem[{\citenamefont{Kao et~al.}(2012)\citenamefont{Kao, Verhulst,
  Vandenberghe, Soree, Groeseneken, and De~Meyer}}]{kao2012}
\bibinfo{author}{\bibfnamefont{K.-H.} \bibnamefont{Kao}},
  \bibinfo{author}{\bibfnamefont{A.~S.} \bibnamefont{Verhulst}},
  \bibinfo{author}{\bibfnamefont{W.~G.} \bibnamefont{Vandenberghe}},
  \bibinfo{author}{\bibfnamefont{B.}~\bibnamefont{Soree}},
  \bibinfo{author}{\bibfnamefont{G.}~\bibnamefont{Groeseneken}},
  \bibnamefont{and} \bibinfo{author}{\bibfnamefont{K.}~\bibnamefont{De~Meyer}},
  \bibinfo{journal}{Electron Devices, IEEE Transactions on}
  \textbf{\bibinfo{volume}{59}}, \bibinfo{pages}{292} (\bibinfo{year}{2012}).

\bibitem[{\citenamefont{Lu et~al.}(2015)\citenamefont{Lu, Esseni, and
  Seabaugh}}]{lu2015}
\bibinfo{author}{\bibfnamefont{H.}~\bibnamefont{Lu}},
  \bibinfo{author}{\bibfnamefont{D.}~\bibnamefont{Esseni}}, \bibnamefont{and}
  \bibinfo{author}{\bibfnamefont{A.}~\bibnamefont{Seabaugh}},
  \bibinfo{journal}{Solid-State Electronics} \textbf{\bibinfo{volume}{108}},
  \bibinfo{pages}{110} (\bibinfo{year}{2015}).

\bibitem[{\citenamefont{Pankove}(1965)}]{pankove1965}
\bibinfo{author}{\bibfnamefont{J.}~\bibnamefont{Pankove}},
  \bibinfo{journal}{Physical Review} \textbf{\bibinfo{volume}{140}},
  \bibinfo{pages}{A2059} (\bibinfo{year}{1965}).

\bibitem[{\citenamefont{Urbach}(1953)}]{urbach1953}
\bibinfo{author}{\bibfnamefont{F.}~\bibnamefont{Urbach}},
  \bibinfo{journal}{Physical Review} \textbf{\bibinfo{volume}{92}},
  \bibinfo{pages}{1324} (\bibinfo{year}{1953}).

\bibitem[{\citenamefont{Subashiev et~al.}(2010)\citenamefont{Subashiev,
  Semyonov, Chen, and Luryi}}]{subashiev2010}
\bibinfo{author}{\bibfnamefont{A.~V.} \bibnamefont{Subashiev}},
  \bibinfo{author}{\bibfnamefont{O.}~\bibnamefont{Semyonov}},
  \bibinfo{author}{\bibfnamefont{Z.}~\bibnamefont{Chen}}, \bibnamefont{and}
  \bibinfo{author}{\bibfnamefont{S.}~\bibnamefont{Luryi}},
  \bibinfo{journal}{Applied Physics Letters} \textbf{\bibinfo{volume}{97}},
  \bibinfo{pages}{181914} (\bibinfo{year}{2010}).

\bibitem[{\citenamefont{Johnson and Tiedje}(1995)}]{johnson1995}
\bibinfo{author}{\bibfnamefont{S.}~\bibnamefont{Johnson}} \bibnamefont{and}
  \bibinfo{author}{\bibfnamefont{T.}~\bibnamefont{Tiedje}},
  \bibinfo{journal}{Journal of applied physics} \textbf{\bibinfo{volume}{78}},
  \bibinfo{pages}{5609} (\bibinfo{year}{1995}).

\bibitem[{\citenamefont{Greeff and Glyde}(1995)}]{greeff1995}
\bibinfo{author}{\bibfnamefont{C.}~\bibnamefont{Greeff}} \bibnamefont{and}
  \bibinfo{author}{\bibfnamefont{H.}~\bibnamefont{Glyde}},
  \bibinfo{journal}{Physical Review B} \textbf{\bibinfo{volume}{51}},
  \bibinfo{pages}{1778} (\bibinfo{year}{1995}).

\bibitem[{\citenamefont{Brammertz et~al.}(2009)\citenamefont{Brammertz, Lin,
  Caymax, Meuris, Heyns, and Passlack}}]{brammertz2009}
\bibinfo{author}{\bibfnamefont{G.}~\bibnamefont{Brammertz}},
  \bibinfo{author}{\bibfnamefont{H.-C.} \bibnamefont{Lin}},
  \bibinfo{author}{\bibfnamefont{M.}~\bibnamefont{Caymax}},
  \bibinfo{author}{\bibfnamefont{M.}~\bibnamefont{Meuris}},
  \bibinfo{author}{\bibfnamefont{M.}~\bibnamefont{Heyns}}, \bibnamefont{and}
  \bibinfo{author}{\bibfnamefont{M.}~\bibnamefont{Passlack}},
  \bibinfo{journal}{Applied Physics Letters} \textbf{\bibinfo{volume}{95}},
  \bibinfo{pages}{2109} (\bibinfo{year}{2009}).

\bibitem[{\citenamefont{Selberherr}(2012)}]{selberherr2012}
\bibinfo{author}{\bibfnamefont{S.}~\bibnamefont{Selberherr}},
  \emph{\bibinfo{title}{Analysis and simulation of semiconductor devices}}
  (\bibinfo{publisher}{Springer Science \& Business Media},
  \bibinfo{year}{2012}).

\bibitem[{\citenamefont{Mohata et~al.}(2012)\citenamefont{Mohata, Rajamohanan,
  Mayer, Hudait, Fastenau, Lubyshev, Liu, and Datta}}]{mohata2012}
\bibinfo{author}{\bibfnamefont{D.}~\bibnamefont{Mohata}},
  \bibinfo{author}{\bibfnamefont{B.}~\bibnamefont{Rajamohanan}},
  \bibinfo{author}{\bibfnamefont{T.}~\bibnamefont{Mayer}},
  \bibinfo{author}{\bibfnamefont{M.}~\bibnamefont{Hudait}},
  \bibinfo{author}{\bibfnamefont{J.}~\bibnamefont{Fastenau}},
  \bibinfo{author}{\bibfnamefont{D.}~\bibnamefont{Lubyshev}},
  \bibinfo{author}{\bibfnamefont{A.~W.} \bibnamefont{Liu}}, \bibnamefont{and}
  \bibinfo{author}{\bibfnamefont{S.}~\bibnamefont{Datta}},
  \bibinfo{journal}{Electron Device Letters, IEEE}
  \textbf{\bibinfo{volume}{33}}, \bibinfo{pages}{1568} (\bibinfo{year}{2012}).

\bibitem[{\citenamefont{Pandey et~al.}(2015)\citenamefont{Pandey, Madan, Liu,
  Chobpattana, Barth, Rajamohanan, Hollander, Clark, Wang, Kim
  et~al.}}]{pandey2015}
\bibinfo{author}{\bibfnamefont{R.}~\bibnamefont{Pandey}},
  \bibinfo{author}{\bibfnamefont{H.}~\bibnamefont{Madan}},
  \bibinfo{author}{\bibfnamefont{H.}~\bibnamefont{Liu}},
  \bibinfo{author}{\bibfnamefont{V.}~\bibnamefont{Chobpattana}},
  \bibinfo{author}{\bibfnamefont{M.}~\bibnamefont{Barth}},
  \bibinfo{author}{\bibfnamefont{B.}~\bibnamefont{Rajamohanan}},
  \bibinfo{author}{\bibfnamefont{M.}~\bibnamefont{Hollander}},
  \bibinfo{author}{\bibfnamefont{T.}~\bibnamefont{Clark}},
  \bibinfo{author}{\bibfnamefont{K.}~\bibnamefont{Wang}},
  \bibinfo{author}{\bibfnamefont{J.-H.} \bibnamefont{Kim}},
  \bibnamefont{et~al.}, in \emph{\bibinfo{booktitle}{VLSI Technology (VLSI
  Technology), 2015 Symposium on}} (\bibinfo{organization}{IEEE},
  \bibinfo{year}{2015}), pp. \bibinfo{pages}{T206--T207}.

\bibitem[{\citenamefont{Mookerjea et~al.}(2009)\citenamefont{Mookerjea, Mohata,
  Krishnan, Singh, Vallett, Ali, Mayer, Narayanan, Schlom, Liu
  et~al.}}]{mookerjea2009}
\bibinfo{author}{\bibfnamefont{S.}~\bibnamefont{Mookerjea}},
  \bibinfo{author}{\bibfnamefont{D.}~\bibnamefont{Mohata}},
  \bibinfo{author}{\bibfnamefont{R.}~\bibnamefont{Krishnan}},
  \bibinfo{author}{\bibfnamefont{J.}~\bibnamefont{Singh}},
  \bibinfo{author}{\bibfnamefont{A.}~\bibnamefont{Vallett}},
  \bibinfo{author}{\bibfnamefont{A.}~\bibnamefont{Ali}},
  \bibinfo{author}{\bibfnamefont{T.}~\bibnamefont{Mayer}},
  \bibinfo{author}{\bibfnamefont{V.}~\bibnamefont{Narayanan}},
  \bibinfo{author}{\bibfnamefont{D.}~\bibnamefont{Schlom}},
  \bibinfo{author}{\bibfnamefont{A.}~\bibnamefont{Liu}}, \bibnamefont{et~al.},
  in \emph{\bibinfo{booktitle}{Electron Devices Meeting (IEDM), 2009 IEEE
  International}} (\bibinfo{organization}{IEEE}, \bibinfo{year}{2009}), pp.
  \bibinfo{pages}{1--3}.

\bibitem[{\citenamefont{Noguchi et~al.}(2013)\citenamefont{Noguchi, Kim,
  Yokoyama, Ji, Ichikawa, Osada, Hata, Takenaka, and Takagi}}]{noguchi2013}
\bibinfo{author}{\bibfnamefont{M.}~\bibnamefont{Noguchi}},
  \bibinfo{author}{\bibfnamefont{S.}~\bibnamefont{Kim}},
  \bibinfo{author}{\bibfnamefont{M.}~\bibnamefont{Yokoyama}},
  \bibinfo{author}{\bibfnamefont{S.}~\bibnamefont{Ji}},
  \bibinfo{author}{\bibfnamefont{O.}~\bibnamefont{Ichikawa}},
  \bibinfo{author}{\bibfnamefont{T.}~\bibnamefont{Osada}},
  \bibinfo{author}{\bibfnamefont{M.}~\bibnamefont{Hata}},
  \bibinfo{author}{\bibfnamefont{M.}~\bibnamefont{Takenaka}}, \bibnamefont{and}
  \bibinfo{author}{\bibfnamefont{S.}~\bibnamefont{Takagi}}, in
  \emph{\bibinfo{booktitle}{Electron Devices Meeting (IEDM), 2013 IEEE
  International}} (\bibinfo{organization}{IEEE}, \bibinfo{year}{2013}), pp.
  \bibinfo{pages}{28--1}.

\end{thebibliography}
\end{document}